\pdfoutput=1
\documentclass[journal]{IEEEtran}

\usepackage{amsmath}
\usepackage{amsfonts}
\usepackage{amssymb}
\usepackage{amsthm}
\usepackage{graphicx}
\usepackage{url}
\usepackage{algorithm2e}
\usepackage{color}

\newcommand{\rar}{\rightarrow}
\newcommand{\lar}{\leftarrow}

\newcommand{\mbb}{\mathbb}
\newcommand{\mbf}{\mathbf}

\definecolor{cdms-red}{rgb}{0.85,0.07,0.23}
\definecolor{cdms-blue}{rgb}{0.24,0.12,0.88} 
\definecolor{cdms-green}{rgb}{0.24,0.66,0.37} 
\definecolor{cdms-purple}{rgb}{0.63,0.08,0.46} 

\begin{document}
\title{Revisiting the Age of Enlightenment from a \\ Collective Decision Making Systems Perspective}

\author{Marko~A.~Rodriguez and~Jennifer~H.~Watkins \\ Los Alamos National Laboratory \\ Los Alamos, New Mexico 87545 %
\thanks{M.A. Rodriguez is with T-5/Center for Nonlinear Studies, Los Alamos National Laboratory, Los Alamos, NM 87545 USA e-mail: marko@lanl.gov.}%
\thanks{Jennifer H. Watkins is with the International and Applied Technology Group, Los Alamos National Laboratory, Los Alamos, NM 87545 USA e-mail: jhw@lanl.gov.}%
\thanks{This research was conducted by the Collective Decision Making Systems (\textbf{\textcolor{cdms-red}{C}\textcolor{cdms-blue}{D}\textcolor{cdms-green}{M}\textcolor{cdms-purple}{S}}) project at the Los Alamos National Laboratory (http://cdms.lanl.gov).}
\thanks{Rodriguez, M.A., Watkins, J.H., ``Revisiting the Age of Enlightenment from a Collective Decision Making Systems Perspective," First Monday, volume 14, number 8, ISSN:1396-0466, LA-UR-09-00324, University of Illinois at Chicago Library, August 2009.}}

%\markboth{Los Alamos National Laboratory Technical Report}
%{Rodriguez and Watkins: Revisiting the Age of Enlightenment}

\maketitle

\begin{abstract}
The ideals of the eighteenth century's Age of Enlightenment are the foundation of modern democracies. The era was characterized by thinkers who promoted progressive social reforms that opposed the long-established aristocracies and monarchies of the time. Prominent examples of such reforms include the establishment of inalienable human rights, self-governing republics, and market capitalism. Twenty-first century democratic nations can benefit from revisiting the systems developed during the Enlightenment and reframing them within the techno-social context of the Information Age. This article explores the application of social algorithms that make use of Thomas Paine's (English: 1737--1809) representatives, Adam Smith's (Scottish: 1723--1790) self-interested actors, and Marquis de Condorcet's (French: 1743--1794) optimal decision making groups. It is posited that technology-enabled social algorithms can better realize the ideals articulated during the Enlightenment.
\end{abstract}


\IEEEpeerreviewmaketitle

\section{Introduction}

Eighteenth century Europe is referred to as The Age of Enlightenment, a period when prominent thinkers began to question traditional forms of authority and power and the moral standards that supported these forms. One of the most significant and enduring contributions of the time was the notion that a government's existence should be predicated on protecting and supporting the natural, immutable rights of its citizens. Among these rights are the right to self-governance, autonomy of thought, and equality. The inherent virtue of these ideas forced many European nations to relinquish time-honored aristocratic and monarchic systems. Moreover, it was the philosophy of the Enlightenment that inspired the formalization of a governing structure that would define a new nation: the United States of America.  

Natural rights exposed during the Enlightenment are immutable. That is, they are rights not granted by the government, but instead are rights inherent to man. However, the systems that maintain and support these rights merit no such permanence. While modern democratic governments strive to achieve the ideals of the Enlightenment, it is put forth that governments can better serve them by making greater use of the technological advances of the present day Information Age. The technological infrastructure that now supports modern nations removes the physical restrictions that dictated many of the design choices of these early government architects. As such, many of today's government structures are remnants of the technological constraints of the eighteenth century. Modern nations have an obligation to improve their systems so as to better ensure the fulfillment of the rights of man. Inscribed at the Jefferson Memorial is this statement by Thomas Jefferson (American: 1743--1826), another thinker of the Enlightenment: ``[...] institutions must go hand in hand with the progress of the human mind. As that becomes more developed, more enlightened [...] institutions must advance also to keep pace with the times." To move in this direction, the principle of citizen representation as articulated by Thomas Paine (English: 1737--1809) and the principle of competitive actors for the common good as articulated by Adam Smith (Scottish: 1723--1790) are considered from a techno-social, collective decision making systems perspective. Moreover, the rationale for these principles can be understood within the mathematical formulations of Marquis de Condorcet's (French: 1743--1794) requirements for optimal decision making. 

\section{The Condorcet Jury Theorem: Ensuring Optimal Decision Making}

\begin{figure}[h]
	\begin{center}
		\includegraphics[width=0.25\textwidth]{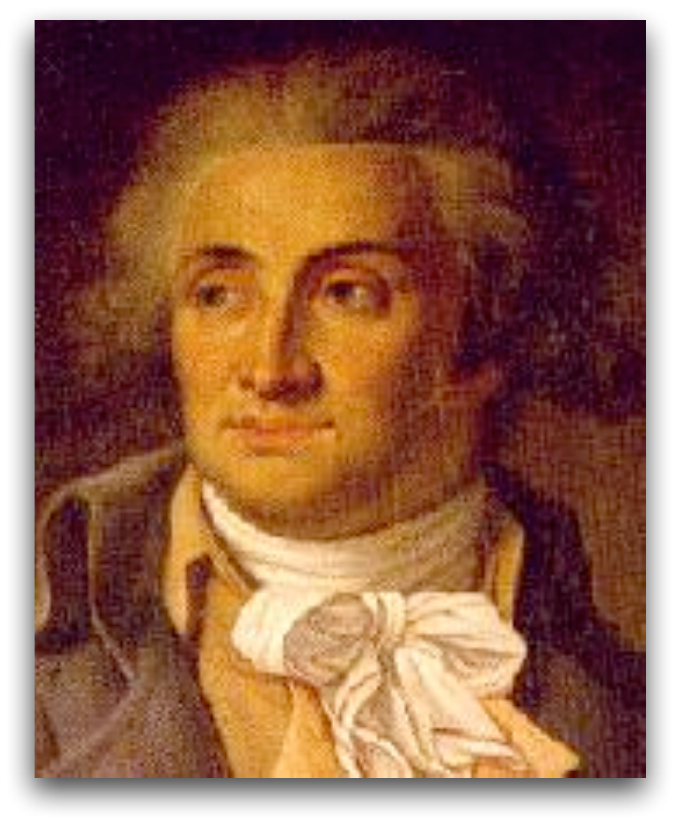}	
		\caption{\label{fig:condorcet-portrait} A portrait of Marquis de Condorcet. This is a public domain photograph courtesy of Wikimedia Commons.}
	\end{center}
\end{figure}

\begin{figure}[h]
	\begin{center}
		\includegraphics[width=0.485\textwidth]{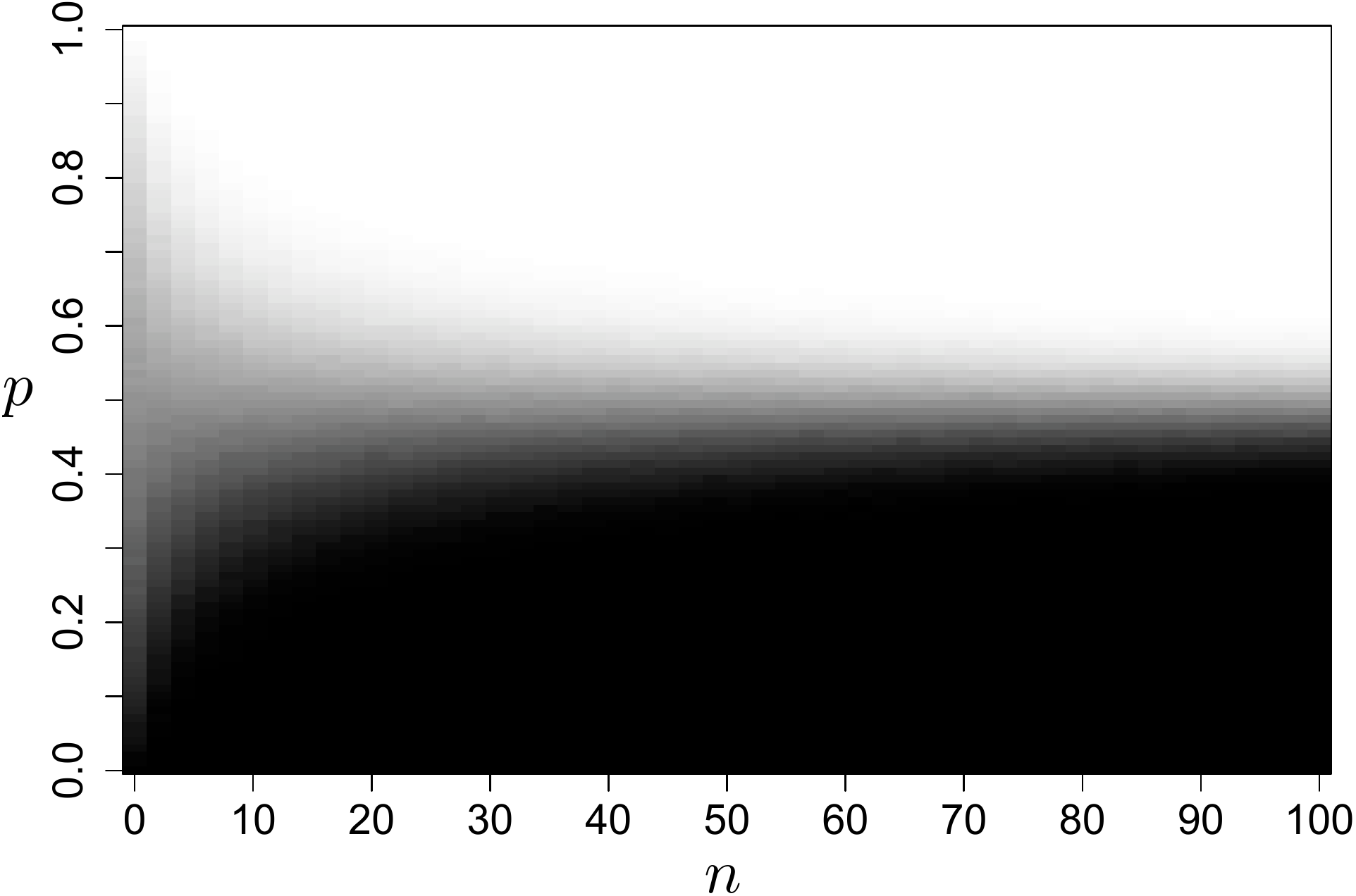}	
		\caption{\label{fig:condorcet} The relationship between $p \in [0,1]$ and $n \in (1,2,\ldots,100)$ according to the Condorcet jury theorem model. Darker values represent a lower probability of a majority vote rendering the best decision and the lighter values represent a higher probability of a majority vote rendering the best decision.}
	\end{center}
\end{figure}

Marquis de Condorcet (portrayed in Figure \ref{fig:condorcet-portrait}) ardently supported equal rights and free and universal public education. These ideals were driven as much by his ethics as they were by his mathematical investigations into the requirements for optimal decision making.
%%%
%%%
One of his most famous results is the Condorcet statement and its associated theorem. In his 1785 \textit{Essai sur l'Application de l'Analyse aux Probabilit\'{e}s des Decisions prises \`{a} la Pluralit\'{e} des Voix} (english translation: \textit{Essay on the Application of Analysis to the Probability of Majority Decisions}), Condorcet states that when a group of ``enlightened" decision makers chooses between two options under a majority rule, then as the size of the decision making population tends toward infinity, it becomes a certainty that the best choice is rendered \cite{condorcet:theorem1776}. The first statistical proof of this statement is the Condorcet jury theorem. The model is expressed as follows. Imagine there exists $n$ independent decision makers and each decision maker has a probability $p \in [0,1]$ of choosing the best of two options in a decision. If $p > 0.5$, meaning that each individual decision maker is enlightened, and as $n \rar \infty$, the probability of a majority vote outcome rendering the best decision approaches certainty at $1.0$. This is known as the ``light side" of the Condorcet jury theorem. The ``dark side" of the theorem states that if $p < 0.5$ and as $n \rar \infty$, the probability of a majority vote outcome rendering the best decision approaches $0.0$. Figure \ref{fig:condorcet} plots the relationship between $p$ and $n$, where the gray scale values denote a range from 100\% probability of the group rendering the best decision (white) to a 0\% probability (black).

The Condorcet jury theorem is one of the original formal justifications for the application of democratic principles to government. While the theorem does not reveal any startling conditions for a successful democracy, it does distill the necessary conditions to two variables (under simple assumptions). If a decision making group has a large $n$ and a $p > 0.5$, then the group is increasing its chances of optimal decision making. Unfortunately, the theorem does not suggest a means to achieve these conditions, though in practice many mechanisms do exist that strive to meet them. For instance, democracies do not rely on a single decision maker, but instead use senates, parliaments, and referendums to increase the size of their voting population. Moreover, for general elections, equal voting rights facilitate large citizen participation. Furthermore, democratic nations tend to promote universal public education so as to ensure that competent leaders are chosen from and by an enlightened populace. It is noted that the practices employed by democratic nations to ensure competent decision making are implementation choices, and a society must not value the implementation of its government. Tradition must be forgone if another implementation would serve better. Implementations of government should be altered and amended so as to better realize the ideals of the nation.

Technology-enabled social algorithms may provide a means by which to reliably achieve the conditions of the ``light" side of the Condorcet jury theorem, thus ensuring optimal decision making. Furthermore, modern algorithms have the potential to do so in a manner that better honors the right of each citizen to participate in government decision making as such algorithms are not constrained by eighteenth century technology. Present day social algorithms, in the form of information retrieval and recommendation services, already contribute significantly to the augmentation of human and social intelligence \cite{webci:watkins2007}. In line with these developments, this article presents two social algorithms that show promise as mechanisms for governance-based collective decision making. One algorithm exaggerates Thomas Paine's citizen representation in order to accurately simulate the behavior of a large decision making population ($n \rar \infty$), and the other employs Adam Smith's market philosophy to induce participation by the enlightened within that population ($p \rar 1$). Both algorithms utilize the Condorcet jury theorem to the society's advantage.

\section{Dynamically Distributed Democracy: Simulating a Large Decision Making Population}

\begin{figure}[h]	
	\begin{center}
		\includegraphics[width=0.25\textwidth]{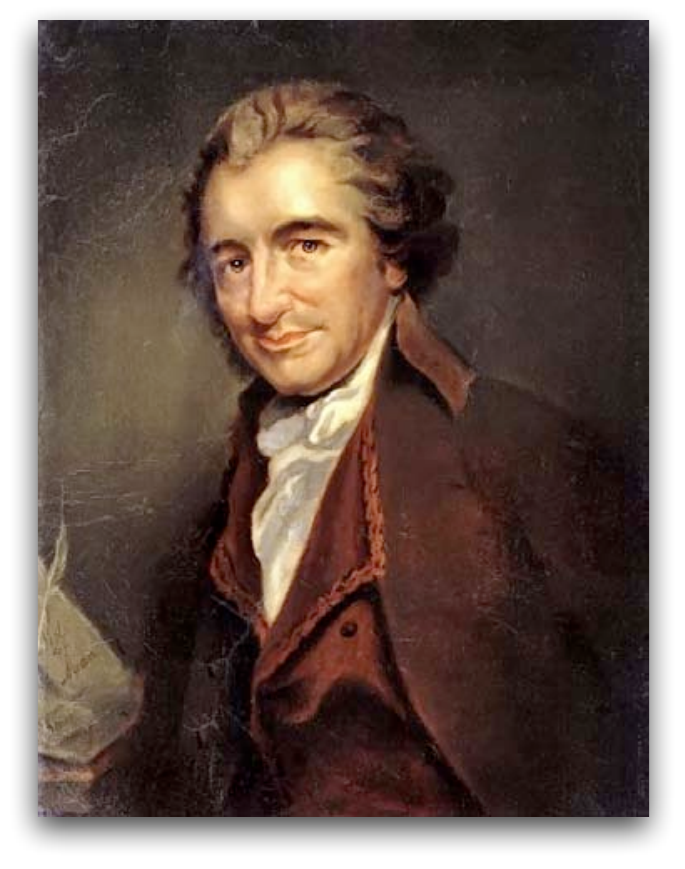}	
		\caption{\label{fig:paine-portrait} An oil portrait of Thomas Paine painted by Auguste Milli\`{e}re in 1880. This is a public domain photograph courtesy of Wikimedia Commons.}
	\end{center}
\end{figure}

Thomas Paine (portrayed in Figure \ref{fig:paine-portrait}) was born in England, but in his middle years, he relocated to America on the recommendation of Benjamin Franklin. It was in America, in the time leading up to the American Revolution, that his enlightened ideals were well received. In 1776, the year in which the Declaration of Independence was written, Thomas Paine wrote a widely distributed pamphlet entitled \textit{Common Sense} which outlined the values of a democratic society \cite{paine:common1776}. This pamphlet discussed the equality of man and the necessity for all those at stake to partake in the decision making processes of the group. As a formal justification of this value, the Condorcet jury theorem would hold that a direct democracy would be the most likely democracy to yield optimal decisions as the voting population is the largest it can possibly be for a nation. In practice, the desire for a direct democracy is tempered by the tremendous burden that constant voting would impinge on citizens (not to mention the logistical problems such a model would incur within present day voting infrastructures). For this reason, representation is required. Thomas Paine states that when populations are small ``some convenient tree will afford them a State house", but as the population increases it becomes a necessity for representatives to act on behalf of their constituents. Moreover, the central tenet of political representation is that representatives ``act in the same manner as the whole body would act were they present." The remainder of this section presents a social algorithm that simulates the manner in which the whole population would act without requiring pre-elected, long-standing representatives.

Assuming a two-option majority rule, an individual citizen's judgement can be placed along a continuum between the two options such that the  ``political tendency" of citizen $i$ is denoted $\mbf{x}_i \in [0,1]$. For example, given United States politics, a political tendency of $0$ represents a fully Republican perspective, a tendency of $1$ represents a fully Democratic perspective, and a tendency of $0.5$ denotes a moderate. Given this definition, there are two ways to quantify the population as a whole. One way is to calculate the average tendency of all citizens. That is $d^\text{tend} = \frac{1}{n} \sum_{i=1}^{i \leq n} \mbf{x}_i$, where $d^\text{tend} \in [0,1]$ is the collective tendency of the population. Given a uniform distribution of political tendency within $\mbf{x}$, the collective tendency approaches $0.5$ as the size of the population increases toward infinity. The other way to quantify the group is to require that the citizen's tendency be reduced to a binary option (i.e.~a two option vote). If a citizen has a political tendency that is less than $0.5$, then they will vote $0$. For a tendency greater than $0.5$, they will vote $1$. If they have a tendency equal to $0.5$ then a fair coin toss will determine their vote. This majority wins vote is denoted $d^\text{vote} \in \{0,1\}$.

\begin{figure}[h]	
	\begin{center}
		\includegraphics[width=0.485\textwidth]{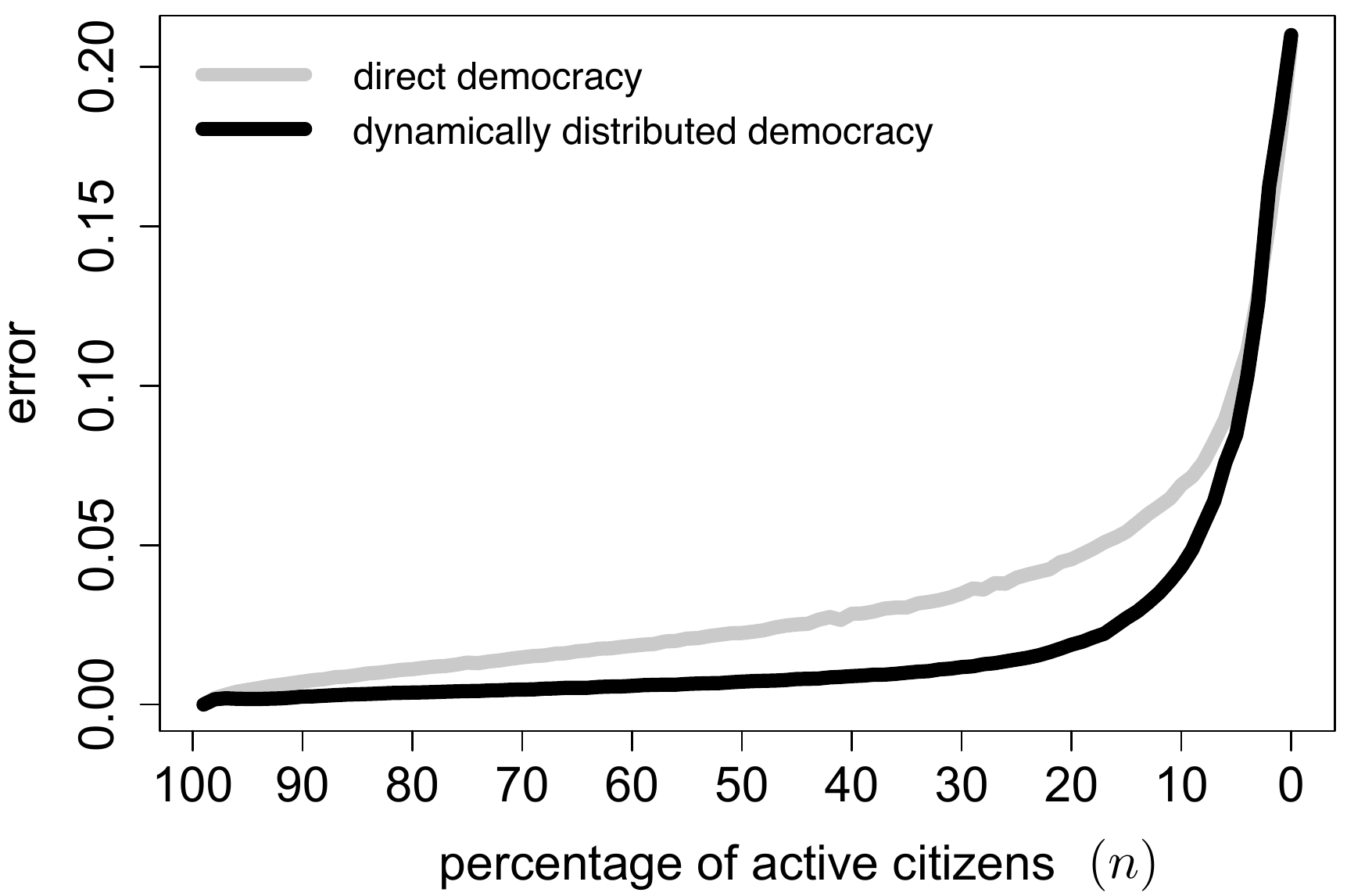}	
		\caption{\label{fig:tendency} The relationship between $k$ and $e_k^\text{tend}$ for direct democracy (gray line) and dynamically distributed democracy (black line). The plot provides the average error over a simulation that was run with 1000 artificially generated networks composed of 100 citizens each.}
	\end{center}
\end{figure}

\begin{figure}[h]	
	\begin{center}
		\includegraphics[width=0.485\textwidth]{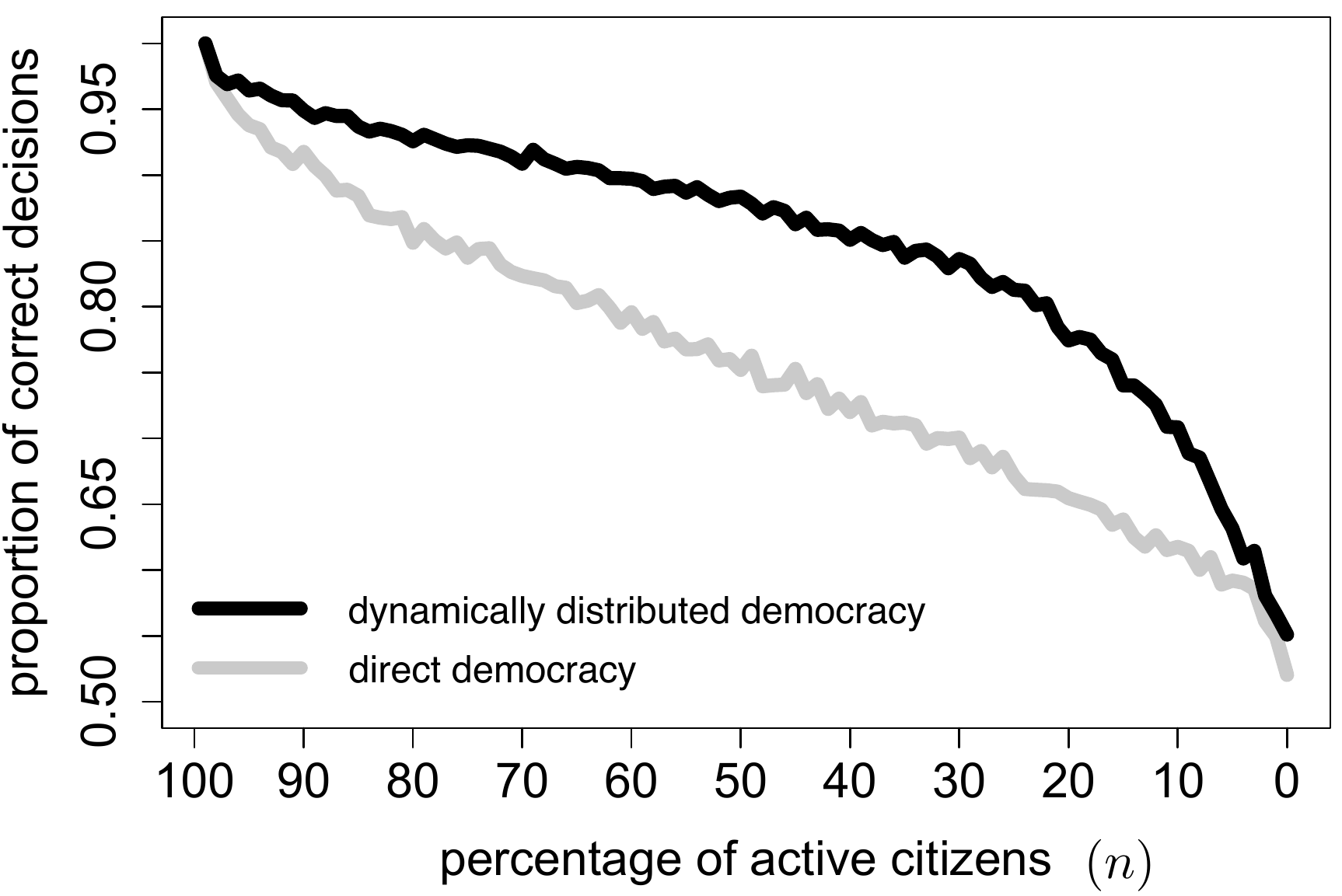}	
		\caption{\label{fig:vote} The relationship between $k$ and $e_k^\text{vote}$ for direct democracy (gray line) and dynamically distributed democracy (black line). The plot provides the proportion of identical, correct decisions over a simulation that was run with 1000 artificially generated networks composed of 100 citizens each.}
	\end{center}
\end{figure}

Imagine a direct democracy in the purest sense, where a raise of hands or a shout of voices is replaced by an Internet architecture and a sophisticated error- and fraud-proof ballot system. All citizens have the potential to vote on any decisions they wish; if they cannot vote on a particular decision for whatever reason, they abstain from participating. In practice, not every decision will be voted on by all $n$ citizens. Citizens will be constrained by time pressures to only participate in those votes in which they are most informed or most passionate. If we assume that all citizens have a tendency, whether they vote or not, how would the collective tendency and collective vote change as citizen participation waned? Let $d_{100}^{\text{tend}} \in [0,1]$ and $d_{100}^{\text{vote}} \in \{0,1\}$ denote the collective tendency and vote given by 100\% participation. Let $d_k^{\text{tend}} \in [0,1]$ and $d_k^{\text{vote}} \in \{0,1\}$ denote the collective tendency and vote if only $k$-percent of the population participates. The error in the collective tendency for $k$-percent participation is calculated as 
%%%
\begin{equation*}
	e_k^\text{tend} = |d_{100}^\text{tend} - d_k^{\text{tend}}|.
\end{equation*}
%%%
The further away the active voters'  collective tendency is from the full population's collective tendency, the higher the error. The gray line in Figure \ref{fig:tendency} plots the relationship between $k$ and $e_k^\text{tend}$. As citizen participation wanes, the ability for the remaining, active participants to reflect the tendency of the whole becomes more difficult. Next, the error in the collective vote is calculated as the proportion of voting outcomes that are different than what a fully participating population would have voted and is denoted $e_k^\text{vote}$. The gray line in Figure \ref{fig:vote} plots the relationship between $k$ and $e_k^\text{vote}$. As participation wanes, the proportion of decisions that differ from what would have occurred given full participation decreases. As with collective tendency, a small active voter population is unable to replicate the voting behavior of the whole.

%%%

Dynamically distributed democracy is a social representation algorithm that provides a means by which any subset of the population can accurately simulate the decision making results of the whole population \cite{ddd:rodriguez2004}. As such, the algorithm reflects the primary tenet of representation as originally outlined by Thomas Paine. The argument for the use of the algorithm as a mechanism for representation goes as follows. Not everyone in a population needs to vote as others in that same population more than likely have a nearly identical political tendency and thus, identical vote. What does need to be recorded is the frequency of that sentiment in the population. If an active, voting citizen is similar in tendency to $10$ non-active citizens, then the active citizen's ballot can be weighted by $10$ to reflect the tendencies of the non-participating citizens. Dynamically distributed democracy accomplishes this weighting through a similarity- or trust-based social network that is used to propagate voting ``power" to active voters so as to mitigate the error incurred by waning citizen participation.

\begin{figure}[h]
	\begin{center}
		\includegraphics[width=0.4\textwidth]{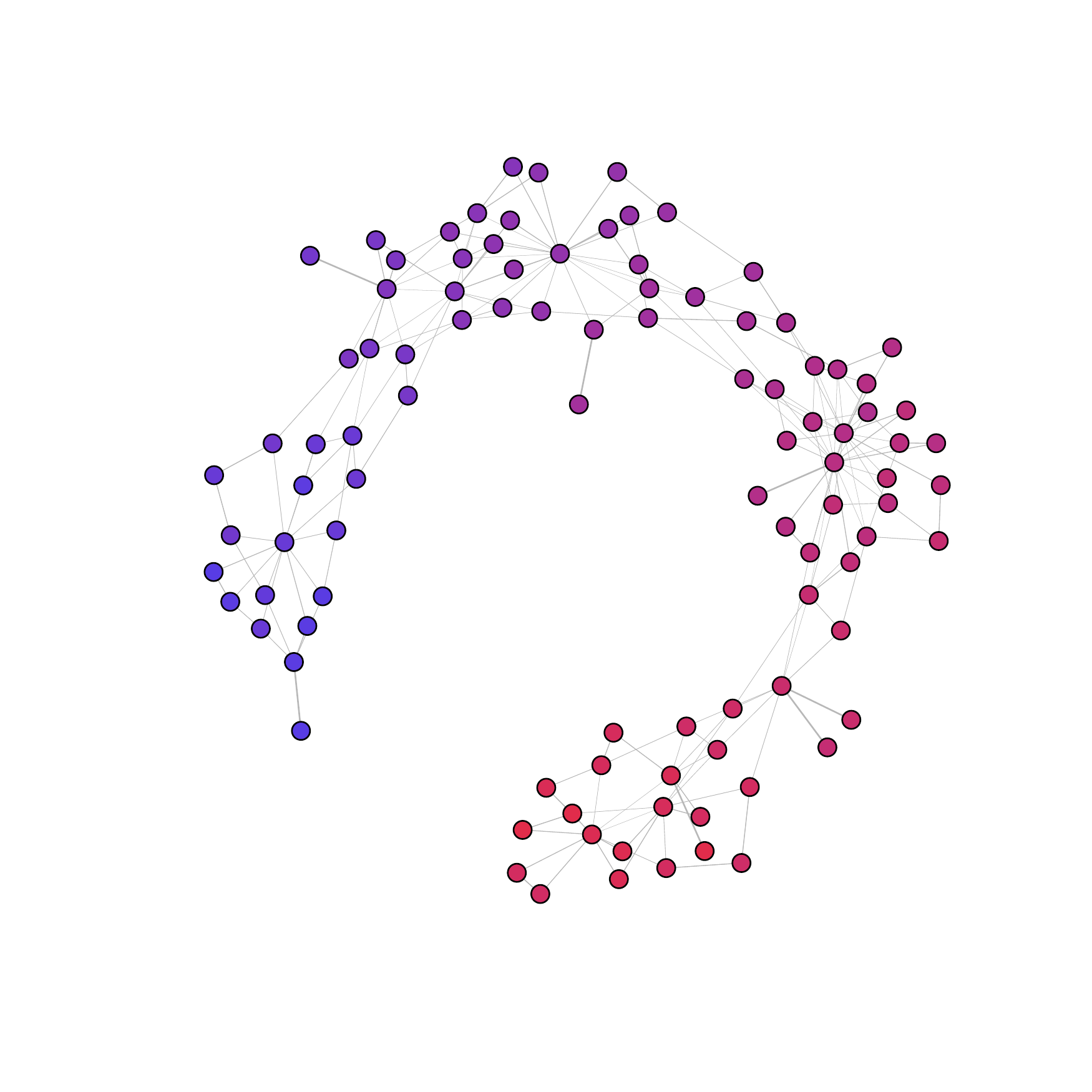}	
		\caption{\label{fig:ddd-network} A visualization of a network of trust links between citizens. Each citizen's color denotes their ``political tendency", where full red is $0$, full blue is $1$, and purple is $0.5$. The layout algorithm chosen is the Fruchterman-Reingold layout.}
	\end{center}
\end{figure}

As previously stated, let $\mbf{x} \in [0,1]^n$ denote the political tendency of each citizen in this population, where $\mbf{x}_i$ is the tendency of citizen $i$ and, for the purpose of simulation, is determined from a uniform distribution. Assume that every citizen in a population of $n$ citizens uses some social network-based system to create links to those individuals that they believe reflect their tendency the best. In practice, these links may point to a close friend, a relative, or some public figure whose political tendencies resonate with the individual. In other words, representatives are any citizens, not political candidates that serve in public office. Let $\mbf{A} \in [0,1]^{n \times n}$ denote the link matrix representing the network, where the weight of an edge, for the purpose of simulation, is denoted
%%%
\begin{equation*}
	\mbf{A}_{i,j} =
		\begin{cases}
			1 - \left|\mbf{x}_i - \mbf{x}_j\right| & \text{if link exists}  \\
			0 & \text{otherwise}.
		\end{cases}
\end{equation*}
%%%
In words, if two linked citizens are identical in their political tendency, then the strength of the link is $1.0$. If their tendencies are completely opposing, then their trust (and the strength of the link) is $0.0$. Note that a preferential attachment network growth algorithm is used to generate a degree distribution that is reflective of typical social networks ``in the wild" (i.e.~scale-free properties). Moreover, an assortativity parameter is used to bias the connections in the network towards citizens with similar tendencies. The assumption here is that given a system of this nature, it is more likely for citizens to create links to similar-minded individuals than to those whose opinions are quite different. The resultant link matrix $\mbf{A}$ is then normalized to be row stochastic in order to generate a probability distribution over the weights of the outgoing edges of a citizen. Figure \ref{fig:ddd-network} presents an example of an $n=100$ artificially generated trust-based social network, where red denotes a tendency of $0.0$, purple a tendency of $0.5$, and blue a tendency of $1.0$.

Given this social network infrastructure, it is possible to better ensure that the collective tendency and vote is appropriately represented through a weighting of the active, participating population. Every citizen, active or not, is initially provide with $\frac{1}{n}$ ``vote power" and this is represented in the vector $\mbf{\pi} \in \mbb{R}_+^n$, such that the total amount of vote power in the population is $1$. Let $\mbf{y} \in \mbb{R}_+^n$ denote the total amount of vote power that has flowed to each citizen over the course of the algorithm. Finally, $\mbf{a} \in \{0,1\}^n$ denotes whether citizen $i$ is participating ($\mbf{a}_i = 1$) in the current decision making process or not ($\mbf{a}_i = 0$). The values of $\mbf{a}$ are biased by an unfair coin that has probability $k$ of making the citizen an active participant and $1-k$ of making the citizen inactive. The iterative algorithm is presented below, where $\circ$ denotes entry-wise multiplication and $\epsilon \approx 1$.
%%%
\incmargin{1cm}
\restylealgo{boxed}
%\linesnumbered
\dontprintsemicolon
\begin{algorithm}[ht!]
	$\mbf{\pi} \lar 0$\;% \CommentSty{\#set total vote power received to $0$}\;
	\While{$\sum_{i=1}^{i \leq n} \mbf{y}_i < \epsilon$}{
		$\mbf{y} \lar \mbf{y} + (\mbf{\pi} \circ \mbf{a})$\; %\CommentSty{\#for active citizens, set received power to current power}\;
		$\mbf{\pi} \lar \mbf{\pi} \circ (1 - \mbf{a})$\;% \CommentSty{\#for active citizens, set current power to $0$}\;
		$\mbf{\pi} \lar \mbf{A}\mbf{\pi}$\;% \CommentSty{\#propagate current power to network neighbors}\;
	}
	%\caption{Dynamically Distributed Democracy algorithm\label{alg:ddd}}
\end{algorithm}
\decmargin{1cm}

In words, active citizens serve as vote power ``sinks" in that once they receive vote power, from themselves or from a neighbor in the network, they do not pass it on. Inactive citizens serve as vote power ``sources" in that they propagate their vote power over the network links to their neighbors iteratively until all (or $\epsilon$) vote power has reached active citizens. At this point, the tendency in the active population is defined as $\delta^\text{tend} = \mbf{x} \cdot \mbf{y}$. Figure \ref{fig:tendency} plots the error incurred using dynamically distributed democracy (black line), where the error is defined as 
%%%
\begin{equation*}
	e_k^\text{tend} = |d_{100}^\text{tend} - \delta^{\text{tend}}_k|.
\end{equation*}
%%%
Next, the collective vote $\delta_k^\text{vote}$ is determined by a weighted majority as dictated by the vote power accumulated by active participants. Figure \ref{fig:vote} plots the proportion of votes that are different from what a fully participating population would have rendered (black line). In essence, if a citizen, for any reason, is unable to participate in a decision making process, then they may abstain from participating knowing that the underlying social network will accurately distribute their vote power to their neighbor or neighbor's neighbor. In this way, representation is dynamic, distributed, and democratic.

Thomas Paine outlines that representatives should maintain ``fidelity to the public" and believes this is accomplished through frequent elections \cite{paine:common1776}. The utilization of an Internet-based social network system affords repeated ``elections" in the form of citizens creating outgoing links to other citizens as they please, when they please, and to whom them please. That is, citizens can dynamically choose representatives who need not be picked from only a handful of candidates. Moreover, if a selected representative falters in their ability to represent a citizen, incoming links can be immediately retracted from them. Such an architecture turns the representative's status from that of elected public official to that of a self-intentioned citizen.

While many countries have political institutions that are set up according to a left, right, and moderate agenda, the individual perspectives of a citizen may be more complex. In many cases, a citizen's political tendency may only be amenable to a multi-dimensional representation. In a multi-relational social network, the links are augmented with labels in order to denote the type of trust one citizen has for another. In this way, voting power propagates over the links in a manner that is biased to the domain of the decision. For example, citizen $i$ may trust citizen $j$ in the domain of ``education" but not in the domain of ``health care". This design has been articulated in \cite{socialgrammar:rodriguez2007}. Supporting systems, including the means by which ballots are proposed and issues are discussed, is presented in \cite{turoff:sdss2002}.  

With the Internet, supporting Web technologies, and dynamically distributed democracy, it is possible to dynamically determine a representative-layer of government that accurately reflects a full direct democracy. In this respect, the larger population helps to ensure, according to the Condorcet jury theorem, that the decisions are either definitely right or definitely wrong. Other technologies can be utilized to induce participation by only those that are more likely than not to choose the optimal decision.

\section{Decision Markets: Incentivizing an Enlightened Majority}

\begin{figure}[h]	
	\begin{center}
		\includegraphics[width=0.25\textwidth]{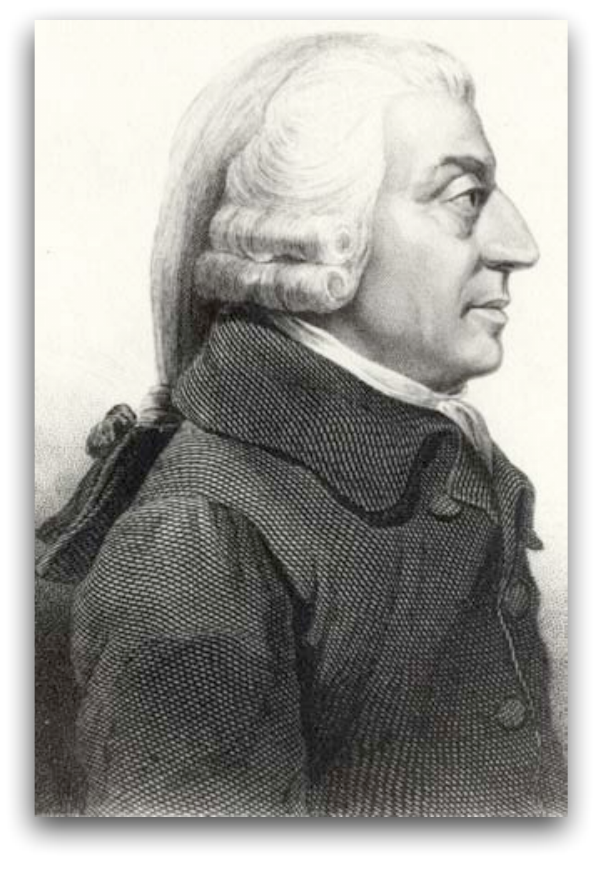}	
		\caption{\label{fig:smith-portrait} An etching of Adam Smith originally created by Cadell and Davies (1811), John Horsburgh (1828), or R.C. Bell (1872). This is a public domain photograph courtesy of Wikimedia Commons.}
	\end{center}
\end{figure} 

Adam Smith (portrayed in Figure \ref{fig:smith-portrait}) was a Scottish moral and economic philosopher who is best known for his two most famous works entitled \textit{The Theory of Moral Sentiments} (1759) and \textit{An Inquiry into the Nature and Causes of the Wealth of Nations} (1776). In the latter work, Adam Smith outlines the economic benefits of a division of labor within a society. Each citizen in the population serves a particular specialized function, and only in the dependency relationships amongst these specialists does an efficient, decentralized economy emerge. With many suppliers and consumers, the production requirements of a society as a whole is difficult to know. Adam Smith appreciated markets for their ability to expose these requirements through the ``natural price" of goods. Moreover, he understood that competition within the market was a necessary driving force guaranteeing an accurate representation of commodity prices. Adam Smith states that when a citizen pursues ``his own interest he frequently promotes that of the society more effectually than when he really intends to promote it"  \cite{wealth:smith1776}.

Market mechanisms are not only useful for determining commodity prices as they can be generally applied to information aggregation and ultimately, to collective decision making. Such markets are called decision markets \cite{hanson:decision}. Similar to a division of labor, the knowledge required to make optimal decisions for a society is dispersed throughout the population. For difficult problems, it is na\"{i}ve to think that a single individual has the requisite knowledge to yield an optimal decision, much like it is na\"{i}ve to think a single merchant will offer the optimal price. A decision market functions because it guarantees a return on investment for quality information. In this respect, a decision market is a tool for attracting a population of knowledgeable citizens much like a commodity market is a tool for attracting knowledgeable speculators. In short, a decision market is a self-selection mechanism that incentivizes participation from those who have knowledge regarding the problem and are confident in their knowledge and discourages participation from others without forbidding it. 

Decision markets reward individuals for buying low and selling high, thus encouraging those who believe they know which way the market will move to contribute their information in the form of the price at which they purchase and sell shares. A decision market differs from commodity markets (such as the New York Stock Exchange) in that stocks represent objective states about the world that can ultimately be determined, but are presently unknown. For example, given the market question ``Will decisions markets be used in U.S. government by the year 2013?", shares of stocks in a ``yes" outcome and in a ``no" outcome are purchased and sold on the market. A high market price for a stock indicates that the collective believes this outcome to be true with a high likelihood. The purpose of the market is to incentivize knowledgeable citizens to contribute to the decision by rewarding them for useful contributions and conversely to inflict a penalty for contributing poor information.

In order to demonstrate the benefits of incentives in decision making, a simulation is provided. Suppose there exists $n$ citizens and a $d$-dimensional ``knowledge space". Each citizen is represented as a point in this space. That is, citizens have different degrees of knowledge in the various dimensions (i.e.~domains) of the space. A citizen's point in this space is generated by a normal distribution with a mean of $p \in [0,1]$ and a variance of $(p(1-p))^2$. Next, there exists an objective truth in this spaced called the environment. For the purpose of simulation, the environment $\mbf{e}$ is the largest valued point in the knowledge space (i.e.~$\mbf{e}_i = 1 : 1 \leq i \leq d$). There also exists a market $\mbf{m}$ which denotes the collective's subjective understanding of the objective environment. For the purpose of simulation, the market  starts as the smallest valued point in the knowledge space (i.e.~$\mbf{m}_i = 0 : 1 \leq i \leq d$). Each citizen participates in the market, moving the market closer or further away from the environment. The closer the market is to the environment, the more accurate the collective decision. There are two markets in the simulation: an incentive-free market and an incentive market. The results of these two markets are compared in order to demonstrate the benefits of using incentives. 

\begin{figure}[h]    
    \begin{center}
        \includegraphics[width=0.35\textwidth]{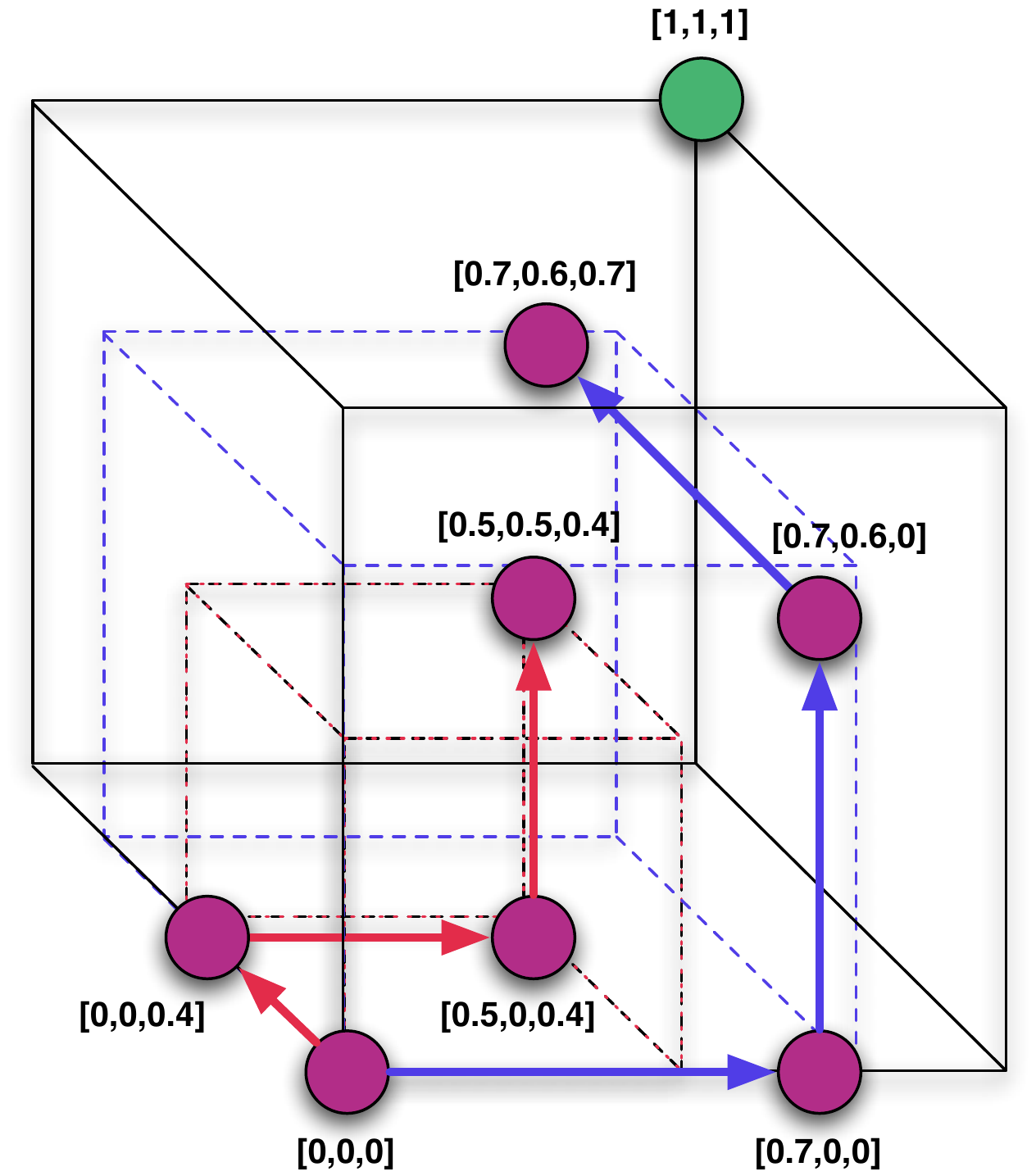}    
        \caption{\label{fig:market-cube}The states of the incentive-free and incentive markets (purple) and the objective state of the environment (green) are diagrammed in a $3$-dimensional knowledge space. There exists two paths: the incentive-free market path (red) and the incentive market path (blue). The dotted cubes denote the range of an incentive-free market (red - $0.5$) and incentive market (blue - $0.75$) for a $p = 0.5$. Refer to the text for a description of the diagrammed market paths.}
    \end{center}
\end{figure}

Before presenting the results of a larger simulation, a small diagrammed example is provided to better elucidate the simulation rules. Figure \ref{fig:market-cube} diagrams a $3$-dimensional knowledge space with both markets (bottom left purple point) and an environment (top right green point). The behavior of the citizens denotes the market paths (red and blue arrows). Also, there exists a $p = 0.5$ population of $3$ citizens, where citizen $\mbf{c}^1 = [0.7,0.5,0.4]$, citizen $\mbf{c}^2 = [0.5,0.6,0.3]$, and citizen $\mbf{c}^3 = [0.3,0.5,0.7]$. At time step $t=0$, both the incentive-free and incentive markets are at $[0,0,0]$. At $t=1$, citizen $\mbf{c}^1$ participates in both markets. In the incentive-free market, citizen $\mbf{c}^1$ has no incentive to contribute his best knowledge and thus, randomly chooses a  dimension in which to move the market. According to the diagram, the citizen's random choice moves the market in the $3^\text{rd}$ dimension by $0.4$. In the incentivized market, citizen $\mbf{c}^1$ chooses the dimension in which he has the most knowledge (i.e.~the dimension with the maximum value). Moreover, a biased coin toss determines whether he participates or not, where $\mbf{c}^1$ has a $70$\% chance of participating in the incentive market. Assuming the coin toss permits it, $\mbf{c}^1$ moves the incentivized market in the $1^\text{st}$ dimension to $0.7$. This process continues in sequence for citizens $\mbf{c}^2$ and $\mbf{c}^3$. Assuming that all citizens participate in both markets, at the end, the incentive-free market is located at point $[0.5,0.5,0.4]$, while the incentive market is located at point $[0.7,0.6,0.7]$. The market error is calculated as the normalized Euclidean distance between the final market position and the environment for a given $p$,
%%%
\begin{equation*}
    e_p^\text{dist} = \frac{1}{\sqrt{d}}\sqrt{\sum_{i=1}^{i \leq d} (\mbf{e}_i - \mbf{m}_i)^2}.
\end{equation*}
%%%
The incentive-free market has an error of $0.287$ and the incentive market has an error of $0.113$. Thus, the incentive market is closest to the environment. There are two distinctions between the markets.  In the incentive-free market, there is no benefit to producing an enlightened solution, so the citizen makes a contribution without comparing his knowledge against the environment. In the incentive market, there are two incentive structures. The first incentive is to participate along the dimension in which the citizen is most knowledgeable. The second incentive is to participate only if the citizen has a satisfactory degree of knowledge. This means that poor information is excluded from the market and that the most valuable knowledge of the citizen is included. 

\begin{figure}[h]    
    \begin{center}
        \includegraphics[width=0.485\textwidth]{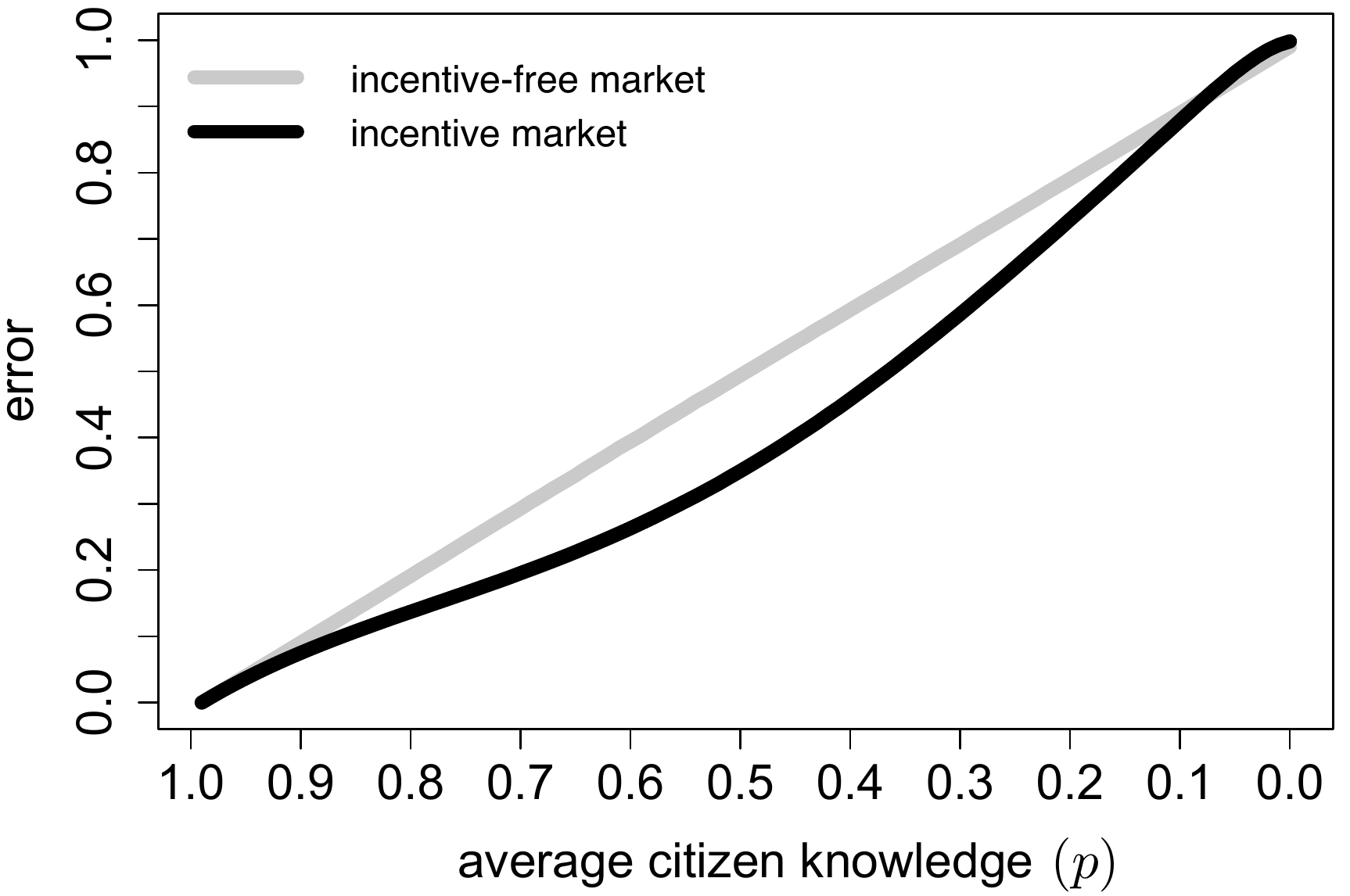}    
        \caption{\label{fig:market} The relationship between $p$ and $e_p^\text{dist}$ for an incentive-free market (gray line) and an incentive market (black line). The plot provides the average error over $1000$ simulations with $d=50$ and $n=1000$.}
    \end{center}
\end{figure}

\begin{figure}[h]    
    \begin{center}
        \includegraphics[width=0.485\textwidth]{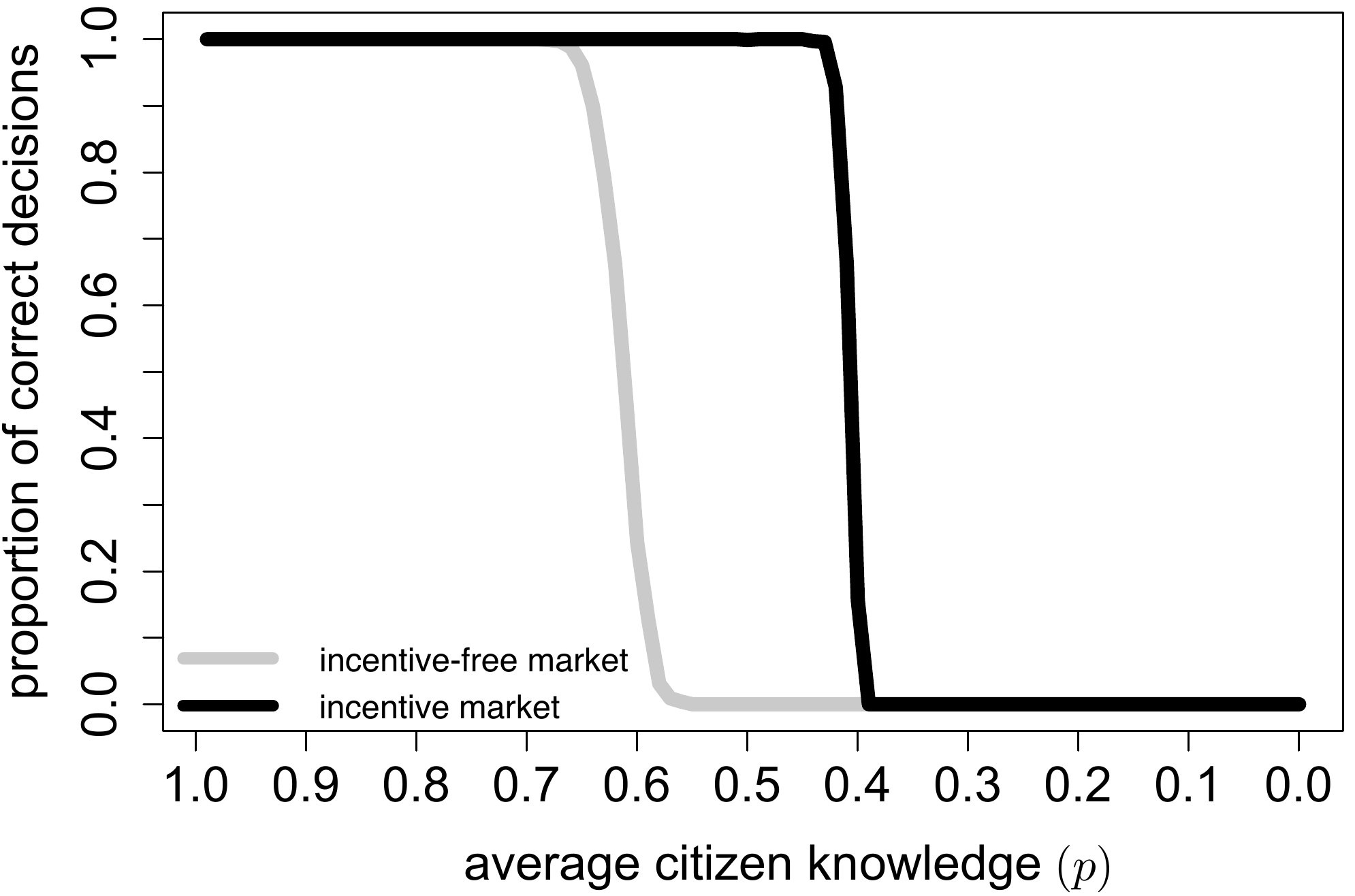}    
        \caption{\label{fig:prediction} The relationship between $p$ and $e_p^\text{deci}$ for an incentive-free market (gray line) and an incentive market (black line). The plot provides the proportion of correct decisions over $1000$ simulations with $d=50$ and $n=1000$.}
    \end{center}
\end{figure}

To demonstrate the effects of an incentive-free and incentive market on a larger population, over various values of $p$, and in a $50$-dimensional knowledge space simulation results are provided. Figure \ref{fig:market} depicts the normalized Euclidean distance error of the incentive-free market (gray line) and the incentive market (black line) for varying $p$. Next, Figure \ref{fig:prediction} provides the proportion of correct collective decisions. A decision is either correct or incorrect. While the market yields a point in $[0,1]^d$, rounding the dimension values of the point to either $1$ or $0$ provides the final decision made by the citizens. For a given $p$, the proportion of times that the market rounds to the environment is the proportion of correct decisions and is denoted $e_p^\text{deci} \in [0,1]$.

It is the principle of self-selection, and therefore citizen choice, that provides the mechanism by which knowledge is aggregated. Choice is manifested in a number of ways. First, citizens choose whether or not to participate in the market at all. This reduces the amount of poor information that enters the market. Second, citizens choose how often to participate. The market, therefore, induces citizens to become more knowledgeable so as to gain from the market. Finally, citizens choose the extent of their participation. If a citizen has knowledge that is not well reflected in the market, suggesting that their knowledge is unique and therefore valuable, the citizen is incentivized to participate more so than if the market closely mimics their knowledge. In decision markets, it is the pricing mechanism of the market that serves the incentivizing role. However, the asset traded in the market need not be money. To maintain the egalitarian nature of self-selection, the market can be based in virtual money with rewards, reputation, or other social inducements as the backing. It has been demonstrated that virtual money is able to preserve the accuracy of decision markets \cite{money:servan2004}.

As presented in the simulation the decisions of a society are multi-dimensional. It is likely that no single citizen has the requisite knowledge in all dimensions to make informed decisions. The ability to reach an optimal decision is dependent on the many dimensions such that ignorance of one dimension may lead to a suboptimal conclusion. The probability parameter of the Condorcet jury theorem model is misleading. It is not through probability that one achieves an optimal decision, but through the careful application of knowledge to the decision. The use of a market is not a guarantee that decision makers have $p > 0.5$. The market is a guarantee that citizen knowledge has been thoughtfully applied to the decision.

\section{Conclusion}

The purpose of a democratic government is to preserve and support the ideals of its population. The ideals established during the Enlightenment are general in nature: life, liberty, and the pursuit of happiness. In articulating these values, the founders of modern democracies provided a moral heritage that remains highly regarded in societies today. However, it should be remembered that it is the ideals that are valuable, not the specific implementation of the systems that protect and support them. If there is another implementation of government that better realizes these ideals, then, by the rights of man, it must be enacted. It was the great thinkers of the eighteenth century Enlightenment who provided the initial governance systems. It is the challenge and the mandate of the Information Age to redesign these governance systems in light of present day technologies.



\begin{thebibliography}{1}
\providecommand{\url}[1]{#1}
\csname url@samestyle\endcsname
\providecommand{\newblock}{\relax}
\providecommand{\bibinfo}[2]{#2}
\providecommand{\BIBentrySTDinterwordspacing}{\spaceskip=0pt\relax}
\providecommand{\BIBentryALTinterwordstretchfactor}{4}
\providecommand{\BIBentryALTinterwordspacing}{\spaceskip=\fontdimen2\font plus
\BIBentryALTinterwordstretchfactor\fontdimen3\font minus
  \fontdimen4\font\relax}
\providecommand{\BIBforeignlanguage}[2]{{%
\expandafter\ifx\csname l@#1\endcsname\relax
\typeout{** WARNING: IEEEtran.bst: No hyphenation pattern has been}%
\typeout{** loaded for the language `#1'. Using the pattern for}%
\typeout{** the default language instead.}%
\else
\language=\csname l@#1\endcsname
\fi
#2}}
\providecommand{\BIBdecl}{\relax}
\BIBdecl

\bibitem{condorcet:theorem1776}
M.~de~Condorcet, ``Essai sur l'application de l'analyse {\'a} la
  probabilit{\'e} des d{\'e}cisions rendues {\'a} la pluralit{\'e} des voix,''
  Paris, France, 1785.

\bibitem{webci:watkins2007}
\BIBentryALTinterwordspacing
J.~H. Watkins and M.~A. Rodriguez, \emph{Evolution of the Web in Artificial
  Intelligence Environments}, ser. Studies in Computational Intelligence.\hskip
  1em plus 0.5em minus 0.4em\relax Berlin, DE: Springer-Verlag, 2008, ch. A
  Survey of Web-Based Collective Decision Making Systems, pp. 245--279.
  [Online]. Available:
  \url{http://repositories.cdlib.org/hcs/WorkingPapers2/JHW2007-1/}
\BIBentrySTDinterwordspacing

\bibitem{paine:common1776}
T.~Paine, ``Common sense,'' American colonies, 1776.

\bibitem{ddd:rodriguez2004}
\BIBentryALTinterwordspacing
M.~A. Rodriguez and D.~J. Steinbock, ``A social network for societal-scale
  decision-making systems,'' in \emph{Proceedingss of the {N}orth {A}merican
  {A}ssociation for {C}omputational {S}ocial and {O}rganizational {S}cience
  {C}onference}, Pittsburgh, PA, 2004. [Online]. Available:
  \url{http://arxiv.org/abs/cs.CY/0412047}
\BIBentrySTDinterwordspacing

\bibitem{socialgrammar:rodriguez2007}
\BIBentryALTinterwordspacing
M.~A. Rodriguez, ``Social decision making with multi-relational networks and
  grammar-based particle swarms,'' in \emph{{P}roceedings of the {H}awaii
  {I}nternational {C}onference on {S}ystems {S}cience}.\hskip 1em plus 0.5em
  minus 0.4em\relax Waikoloa, Hawaii: {IEEE} Computer Society, January 2007,
  pp. 39--49. [Online]. Available: \url{http://arxiv.org/abs/cs.CY/0609034}
\BIBentrySTDinterwordspacing

\bibitem{turoff:sdss2002}
M.~Turrof, S.~R. Hiltz, H.-K. Cho, Z.~Li, and Y.~Wang, ``Social decision
  support system ({SDSS}),'' in \emph{{P}roceedings of the {H}awaii
  {I}nternational {C}onference on {S}ystems {S}cience {H}awaii {I}nternational
  {C}onference on {S}ystems {S}cience}.\hskip 1em plus 0.5em minus 0.4em\relax
  Waikoloa, Hawaii: {IEEE} Computer Society, January 2002, pp. 81--90.

\bibitem{wealth:smith1776}
A.~Smith, \emph{An Inquiry into the Nature and Causes of the Wealth of
  Nations}.\hskip 1em plus 0.5em minus 0.4em\relax London, England: W. Strahan
  and T. Cadell, Londres, 1776.

\bibitem{hanson:decision}
R.~Hanson, ``Decision markets,'' \emph{{IEEE} Intelligent Systems}, vol.~14,
  no.~3, pp. 16--19, May 1999.

\bibitem{money:servan2004}
E.~Servan-Schrieber, J.~Wolfers, D.~M. Pennock, and B.~Galebach, ``Prediction
  markets: Does money matter?'' \emph{Electronic Markets}, vol.~14, no.~3, pp.
  243--251, September 2004.

\end{thebibliography}
\end{document}